\documentstyle[a4,12pt]{article}
\newcommand{\itvec}[1]{\mbox{\boldmath ${#1}$}}
\begin{document}
\noindent

\vspace*{4.7cm}
\noindent\bf
EXACT RESULTS FOR SPIN AND CHARGE DYNAMICS \\
OF ELECTRONS WITH SUPERSYMMETRY \\

\vspace{3em}

\rm
\hspace*{1.5cm}
Y. Kuramoto

\vspace{1em}
\hspace*{1.5cm}
Department of Physics

\hspace*{1.5cm}
Tohoku University

\hspace*{1.5cm}
Sendai 980, JAPAN

\vspace{3em}
\noindent
{\bf INTRODUCTION}

\bigskip
In the presence of strong interactions, spin and charge of electrons behave
differently.  In the half-filled case of the Hubbard model with large on-site
repulsion, for example, charge excitation has a gap but spin excitations remain
gapless.
The dynamics in the large energy scale can be understood in terms of
perturbation theory. However study of the infrared behavior requires more
sophisticated approach like the renormalization group or exact solution.
Concept of the fixed point turns out to be vital in understanding the dynamics
in the infrared limit.
Although starting models have an extreme variety, the number of different fixed
points are very few.  Therefore classification of models according to their
fixed points are useful.

The dynamics characteristic of a fixed point is in general restricted only to a
narrow energy range, and different behavior sets in as the excitation energy is
increased.
Even very near the fixed point, presence of marginally irrelevant operators
causes complication in dynamics and in asymptotic behavior of correlation
functions.  This is also the case for the Hubbard model and the short-range \it
t-J \rm model in one dimension.
Thus it is illuminating to investigate models with simpler but nontrivial
fixed-point behaviors.
Recently a new class of interacting fermion models have been found where the
fixed-point behavior persists to all excitation energies.$^{1-6}$  A
fundamental role in this surprising phenomena is played by the supersymmetry
which is associated with transformation of a fermion into a boson and vice
versa.

In this paper we discuss spin and charge dynamics in
a SU($\nu$) generalization of the one-dimensional \it t-J \rm model with
long-range interactions.
In this model dynamics of spin and charge are independent of each other for all
energies.  Moreover there is a discontinuity in the momentum distribution in
the ground state, in contrast to the power-law singularity which is ubiquitous
in one dimension.  We have ascribed these properties to realization of the
fixed point of free Luttinger liquids for spin and charge.$^{1-3}$
If one takes the high-density limit and freezes the charge degrees of freedom,
the model is reduced to the long-range exchange model proposed by Haldane and
Shastry.$^{7,8}$

Complete integrability of a group of long-range models has been proven by
finding infinite number of conserved quantities.$^{9-13}$
Operator algebra has been very powerful in the proof, but has been less so in
deriving wave functions.
As a complementary, we take in this paper a more primitive approach which can
derive wave functions explicitly.  In our representation one can see clearly
how the independent charge- and spin-current excitations influence the energy
of the system.
Our formalism makes maximum use of rotational invariance of the model as well
as that of the ground-state wave function.  The technique is rather general and
can be applied to a class of related models.$^{14,15}$

We briefly discuss also a supersymmetric continuum model$^{16}$ which
constitutes a fixed point of free non-relativistic SU($\nu$) fermions.  The
unique feature of the latter model is that it is solvable in arbitrary
dimensions.  The model has a close relation to the sigma model.$^{17}$

\vspace{10mm}

\noindent
{\bf SU($\nu$,1) SUPERSYMMETRIC \itvec t-\itvec J MODEL}

\bigskip
The \it t-J \rm model is a standard model for studying strong correlation of
electrons in copper oxides and other related systems.  In this paper we
generalize up and down components of spins to $\nu$ kinds of internal degrees
of freedom, which we still call spin.  In this SU($\nu$) version, the \it t-J
\rm model is defined by
\begin{equation}
 {\cal H} = {\cal P}\sum _{i\neq j}\, [-t_{ij}\sum_\sigma
f_{i\sigma}^\dagger f_{j\sigma}+{1\over 2}J_{ij} (P_{ij}-n_in_j)]{\cal P},
\label{eq:model}
\end{equation}
where $f_{j\sigma}$ represents the annihilation operator of an electron at site
$j$ and spin $\sigma (= 1,\ldots , \nu)$, ${\cal P}$ is a projection operator
to exclude double occupation of any site, and $P_{ij}$ denotes permutation
operator of spins.

As the one-dimensional lattice we consider a ring which consists of $N_L$
(even) lattice points $x_i$ with unit spacing and impose a condition $
t_{ij}=J_{ij}$.
In order to show the presence of supersymmetry in the case of $t_{ij}=J_{ij}$,
it is convenient to introduce a hard-core electron at site $i$ as a composite
particle of a fermion with a creation operator $c_i^\dagger$ and a boson with
$b_{\sigma i}^\dagger$.
We represent the creation operator of a hard-core electron as
$ X^{\sigma 0}_i \equiv b_{\sigma i}^\dagger c^\dagger_i
$.
The spin flipping operator from $\beta$ to $\alpha$ is defined by
$
X^{\alpha\beta}_i=b_{\alpha i}^\dagger b_{\beta i}.
$
We further define
$
X^{00}_i = c_i c_i^\dagger ,
$
which is the projection operator onto the fermionic vacuum. A general
$X$-operator is represented by $X^{ab}$ where $a$ and $b$ denote either 0 or
one of $\sigma $.

The commutation rules of these $X$-operators at a site obeys those of a Lie
superalgebra associated with the supergroup SU($\nu$,1).$^{18}$
Hence the $X$-operators generate, like the spin operators in the case of SU(2),
``superrotation" which mixes $c_i^\dagger$ and $b_{\sigma i}$.
Let us introduce a $(\nu +1)$-dimensional vector operator $\Psi_i$ whose
conjugate $\Psi_i^\dagger$ has components
\begin{equation}
\Psi_i^\dagger =( b_{1i}^\dagger,\cdots, b_{\nu i}^\dagger, h_i^\dagger),
\label{eq:sutri}
\end{equation}
where $h_i^\dagger = c_i$ creates the vacant state.
The superrotation leaves the constraint $I=\Psi_i^\dagger \Psi_i =1$ invariant.
Thus the supersymmetry at the single-site level is merely a representation of
the hard-core constraint of fermions.  However supersymmetry imposes a
nontrivial symmetry for intersite interactions.  We note that an intersite
operator $\Psi_i^\dagger \Psi_j$ is also invariant under the global $(\nu
+1)$-dimensional superrotation which is common to all sites.  Thus if the
intersite interaction in a model is expressible in terms of $\Psi_i^\dagger
\Psi_j$, there remains a global supersymmetry.

Let us introduce a graded permutation operator $\tilde{P}_{ij}$.  It is defined
by
\begin{equation}
\tilde{P}_{ij} = \ :(\Psi_i^\dagger \Psi_j)(\Psi_j^\dagger \Psi_i):\
=-\sum_{a,b} X_i^{ab} X_j^{ba} \theta_b,
\label{eq:perm}
\end{equation}
where $:\cdots :$ indicates the normal ordering of component operators.  If the
ordering makes exchange of two fermion operators, the sign is reversed.  This
operation leads to the second equality where $\theta_b = 1$ if $b=0$, and
$\theta_b =-1$ otherwise.  The presence of the factor $\theta_b$ is
characteristic of the graded permutation.  We note that $\tilde{P}_{ij}$ is
invariant under the superrotation as apparent from the first equality.

The \it t-J \rm model is rewritten as
\begin{equation}
 {\cal H} = \sum _{i\neq j}\, J_{ij}(\tilde{P}_{ij}-1+2X^{00}_i).
\label{eq:susy}
\end{equation}
 From eq.(\ref{eq:susy}), it is clear that the model without the $X^{00}_i$
term has supersymmetry.$^{19}$  This chemical potential term breaks the
SU($\nu$,1) supersymmetry just as the magnetic field breaks the SU(2) symmetry
of the Heisenberg model.  Once the model without the $X^{00}_i$ term is solved,
inclusion of the term is easily done.  We note that the discussion of this
section is general as long as $t_{ij}=J_{ij}$.

\vspace{10mm}
\noindent
{\bf PERMUTATION PROPERTIES OF SU($\nu$) ELECTRONS}

\bigskip
The ground-state wave functions for a group of long-range SU($\nu$) fermion
models have been shown to be of Jastrow form.$^{6,15}$  The derivation uses
elementary but complicated algebra.  The complexity arises when one deals with
permutation of spins.  We note that the permutation property is in fact the
same as that of free SU($\nu$) electrons.  By using this fact, one can bypass
most of the complicated algebra in deriving the ground-state wave function and
the energy of the SU($\nu$,1) \it t-J \rm model.

Let us consider free $N_e$ electrons in the continuum ring with unit radius.
At the ground state the electrons fill the momentum states from zero to the
Fermi wave number for each spin.  This leads to a spin singlet wave function
$\Psi_F \{\theta,\sigma\}$ which is given by the product of Slater
determinants.
Here $\{\theta ,\sigma\}$ denote the set of spatial and spin coordinates with
$0\leq \theta_i <2\pi$ for each particle.  In one dimension, the Slater
determinant is reduced to the Vandermonde determinant and $\Psi_F \{\theta
,\sigma\}$ is given by

$$
\Psi_F \{\theta ,\sigma\} = \exp (iS), $$ $$
S= \sum_{i<j} [\delta(\sigma_i,\sigma_j)\ln (\sin {1\over 2}\theta_{ij}) +
{\pi\over 2} {\rm sgn}
(\sigma_i -\sigma_j)] .
$$

This wave function has a nontrivial permutation property.
To derive this we apply the momentum operator $ p_i = -i\partial /\partial
\theta _i$ to $\Psi_F \{\theta,\sigma\}$.  Then we obtain
\begin{equation}
p_i \Psi_F \{\theta,\sigma\} = -i \sum_{j(\neq i)}
\delta(\sigma_i,\sigma_j)c_{ij} \Psi _F\{\theta,\sigma\} ,
\label{eq:ai}
\end{equation}
where $c_{ij} = 2^{-1}\cot(\theta_{ij}/2)$.
The singlet nature of $\Psi_F \{\theta,\sigma\}$ makes it invariant against any
SU($\nu $) rotation $R$, and obviously $R p_i R^{-1} = p_i$.  Rotational
average of eq.(\ref{eq:ai}) leads to replacement of $\delta(\sigma_i,\sigma_j)$
by
$
f T_{ij}
$
where $T_{ij}$ is a projection of operator to spin symmetric states and the
prefactor $f= 2/(\nu +1)$ corresponds to the fraction of equal-spin pairs to
the total number of symmetric spin states.  In terms of the spin permutation
operator $P_{ij}$, $T_{ij}$ is given by $T_{ij} = (P_{ij}+1)/2$. We then
introduce an operator
$$
a_i = p_i +i \sum_{j(\neq i)} f c_{ij}T_{ij},
$$
which annihilates the ground state for each $i$.
Consequently we obtain
$\sum_i a_i^\dagger a_i \Psi_F =0$.  After some algebra this relation leads to
the following identity:
\begin{equation}
f\sum\,^\prime  c_{ij}c_{i\ell} T_{ij}\Psi_F =
\left[ {1\over 4} (1-f) \sum_{i\neq j}  T_{ij}\sin ^{-2}({1\over
2}\theta_{ij})+E_\nu\right] \Psi _F ,
\label{eq:id}
\end{equation}
$$
E_\nu = -{\nu\over 12} N_\sigma (\nu N_\sigma^2+3N_\sigma+\nu-3) ,
$$
where $N_\sigma$ is the number of electrons for each spin.
The sum with prime runs over coordinate indices $i, j, \ell$ different from
each other.

This identity relates the three-body term in the left-hand side to the two-body
term plus the constant.  Two remarkable features of this identity are worth
noting: First, the identity is valid also for the Gutzwiller-projected wave
function in the one-dimensional lattice, since restriction of $\theta_i$ on the
lattice together with hard-core constraint has no effect on the permutation
property.  Secondly, the identity is also valid for any wave function which is
a product of $\Psi_F$ with a spin-independent function.

As an analogue of the ring system for SU($\nu$) free fermions, one may consider
a linear system where a harmonic potential prevents fermions from going to
infinity.  the wave function of the SU($\nu$) free fermion ground state
$\Phi_F\{ x,\sigma\}$ is easily constructed.  In this case we obtain the
analogous identity which reads
$$
f\sum\,^\prime  \frac{T_{ij}}{(x_i -x_j)(x_i -x_\ell)}\Phi_F =
( 1-f) \sum_{i\neq j}  \frac{T_{ij}}{(x_i-x_j)^2}\Phi _F .
$$
Note that the identity is independent of strength of the harmonic potential,
and there is no constant term.  This identity is useful for exact solution of
the SU($\nu$) Calogero model,$^{14}$ and simplifies the treatment by Vacek et
al.$^{15}$

\vspace{10mm}

\noindent
{\bf WAVE FUNCTION AND ENERGY OF THE GROUND STATE}

\bigskip
 From this point on we restrict ourselves to the case where the interaction has
the long-range form:
$ t_{ij}=J_{ij}=t D(x_i-x_j)^{-2} $
with $D(x)=(N_L/\pi )\sin [\pi x/N_L]$ and assume $t >0$.
For this SU($\nu$,1) \it t-J \rm model Ha and Haldane have given a nice and
detailed account of their calculation$^6$ which is rather involved.  We give in
the following a different treatment which we believe is simpler.
The formal proof that the wave function indeed is the ground state is still
missing, but there are strong evidences in favor of that.$^{1-6}$

The major role in our treatment is played by the identity eq.(\ref{eq:id}) for
the Gutzwiller wave function.  In order to treat the transfer term together
with the hard-core constraint it is convenient to take a fully polarized state,
say in the direction of $\nu$, as the reference state $|N_L \nu\rangle$.  Then
there emerge fermionic holes and hard-core bosons with spin $\sigma \,
(=1,\ldots, \nu -1)$ as particles.
The SU($\nu$) Gutzwiller state $|G\rangle$ is represented by
$$ | G\rangle  =  \sum_{\{x_\sigma\},\{y\}}\Psi_G (\{ x_\sigma\},\{
y\})\prod_{\sigma = 1}^{\nu -1}
\prod_{i \in \{ x_\sigma\}} f_{i\sigma}^\dagger
 f_{i\nu}
\prod_{j\in\{ y\}}f_{j\nu}| N\nu \rangle .
$$
Here $\{ x_\sigma\}$ denotes the set of coordinates for $N_\sigma$ electrons
with spin $\sigma$ and $\{ y\}$ does that of $Q$ holes.  Thus we have $N_L=\nu
N_\sigma+Q$ in the singlet state. In order to remove the degeneracy we choose
$N_\sigma$ odd. The amplitude $\Psi_G (\{ x_\sigma\},\{ y\})$ is given, apart
from a normalization factor, by
\begin{eqnarray}
 \Psi_G (\{ x_\sigma\},\{ y\}) & = & \exp [-i\pi (\sum_{i\sigma} x_{i\sigma}
+\sum_\ell y_\ell )]
\prod_{\sigma} \prod_{i > j } D(x_{i\sigma} -x_{j\sigma} )^2 \nonumber\\
& &  \times \prod_{\alpha \neq \beta}\prod_{i,j } D(x_{i\alpha} -x_{j\beta} )
\prod_\sigma\prod_{i, \ell}D(x_{i\sigma} -y_\ell )
\prod_{\ell >m} D(y_\ell -y_m)
..    \label{eq:g}
\end{eqnarray}

In this representation it easy to apply the transfer operator of electrons with
spin $\nu$, since it is translated into the motion of holes without any effect
on $x_{i\sigma}$.  On the other hand, transfer of electrons with a different
spin has the same effect as that of spin $\nu$ in the singlet state.$^{1}$
Thus the whole effect of the transfer is just $\nu$ times of the hole transfer
in eq.(\ref{eq:g}).
Equivalently, we can take rotational average of the wave function after
calculating the hole transfer, and multiply the result by $\nu$.
As a result of the transfer, we have three-body terms multiplying $\Psi_G$.  At
this state we take the vacant state as a new reference state to make use of the
SU($\nu$) invariance.  Then the three-body term
 turns out to be of the same form as the left-hand side of eq.(\ref{eq:id}).
By using the identity eq.(\ref{eq:id}) we obtain the two-body term which is
just minus of the interaction term in eq.(\ref{eq:model}) and the constant.
The ground-state energy $E_g$ corresponds to minus of this constant and is
calculated as
$$
{E_g\over \pi^2 t} = -\frac{N_L}{6 \nu}(n_e^3-3n_e^2+2\nu n_e)
+{1\over 6 N_L}[(\nu+2)n_e-3\nu] ,
$$
with $n_e = N_e/N_L = \nu N_\sigma /N_L$.
This results agrees with that of ref.6 and reduces to that of ref.1 for $\nu
=2$.  The charge susceptibility $\chi _c$ is calculated from the second
derivative of $E_g$ as $\chi _c = \pi^2 t(1-n_e)/\nu $.

\vspace{10mm}

\noindent
{\bf SPIN AND CHARGE CURRENTS IN THE FIRST QUANTIZATION}

\bigskip
It has been shown that long-range the supersymmetric \it t-J \rm model has
completely separated spin and charge excitations.$^{1-6}$  In this section we
clarify how this separation is related to the particular form of the model.
The single-particle spectrum would be given, if there were no two-body
interactions, by Fourier transform of $J_{ij}$ or $t_{ij}$.  This is calculated
as
\begin{equation}
J(q)=\sum_i J_{ij}\exp (-iq\theta_{ij}) =
(q-\frac{N_L}{2})^2-\frac{1}{12}(N_L^2+2),
\label{eq:fou}
\end{equation}
where $\theta_{ij}=\theta_i-\theta_j=2\pi (x_i-x_j)/N_L$ and $q$ is an integer
with $0\leq q <N_L$.  The unit of energy is so chosen that $J=t=(N_L/\pi)^2/2$.

A remarkable feature of this spectrum is that $J(q)$ is a quadratic function of
$q$.  If one shifts the origin of $q$ to the edge $N_L/2$ of the Brillouin
zone, the spectrum is nearly the same as that of free particles in continuum
space, except for
the presence$^{20}$ of the cut off in $q$.
Using this similarity we can also work conveniently with the first-quantized
representation with the completely polarized state in the direction of $\sigma
=\nu$ as the reference state.
The complete set for the system consists of the product of one-body states
inside the Brillouin zone with proper symmetrization.  In terms of $z_i=\exp
(i\theta _i)$, where $\theta_i$ denotes a coordinate with either a spin other
than $\nu$ or a hole, the one-body states are spanned by monomials $z_i^k$ with
$0\leq k < N_L$.
We now shift the origin of the wave number to the edge $N_L/2$ of the Brillouin
zone.  Within the many-body Hilbert space defined in this way, we may replace
$q-N_L/2$ in eq.(\ref{eq:fou}) by $ -i\partial /\partial\theta$.
Then the Hamiltonian is given by
\begin{equation}
{\cal H} =  \sum_{i=1}^M p_i^2 + {1\over 4}\sum_{i\neq j}\sin ^{-2}({1\over 2}
\theta _{ij})(1 + \tilde{P}_{ij}) + E_M,
\label{eq:fq}
\end{equation}
where $p_i = -i\partial /\partial\theta_i$ and $M=N_L-N_\sigma$ is the number
of holes plus bosons with spins different from $\nu$.  The graded permutation
operator $\tilde{P}_{ij}$ acts now in the space of SU($\nu-1$) spins and holes.
 The constant $E_M$ appears as a result of transformation.

By taking the hole picture in the first quantization, the minus sign in the
hopping combines nicely with the plus sign in $J_{ij}$ in leading to
eq.(\ref{eq:fq}) where $p_i^2$ is common to holes and bosons.
Equation (\ref{eq:fq}) can be regarded as the SU($\nu-1,1$) generalization of
the Sutherland model in continuum space.  Thus all eigenfunctions of the
SU($\nu$,1) \it t-J \rm model can be mapped into those of the SU($\nu-1$,1)
Sutherland model.  However the reverse is not true because of restriction of
wave functions to the Brillouin zone in the \it t-J \rm model.

We introduce a shift $k_\sigma$ in the momentum distribution for each spin in
the original lattice system.
In the presence of small currents for each spin component $\alpha$ or $\beta$,
the energy of the system increases by the amount
$$
\delta E = {\pi\over 2N_L}\sum_{\alpha\beta}v_{\alpha\beta}J_\alpha J_\beta ,
$$
where a current component $J_\alpha = 2k_\sigma$ is an even integer.
The velocity matrix $v_{\alpha\beta}$ is constrained to the two-parameter form
$ v_{\alpha\beta} = \delta _{\alpha\beta}u + v$.
Here the off-diagonal component is independent of spin indices because of the
SU($\nu$) invariance.
By making linear transformation of $J_\sigma$, one can diagonalize the velocity
matrix.  The first eigenvalue corresponds to the charge velocity given by
$u+v$, and the remaining $\nu -1$ ones are spin velocities which are all
degenerate and are given by $v$.  Thus if one can calculate $\delta E$ for two
sets of $\{ J_\sigma \}$, one derives spin and charge velocities.

In the first quantization representation, the operator to generate the current
becomes diagonal in coordinates and is given by
$$
\phi_{\{k\}} \{\theta\}=\exp\left( i \sum_{\sigma\, (\neq\nu)} k_\sigma \sum_i
\theta_{i\sigma}
 - i \sum_j k_\nu \theta _j \right),
$$
where $\theta_{i\sigma}$ refers to a coordinate of a particle with spin
$\sigma$, while $\theta _j$ refers to all particles.
The minus sign with $k_\nu$ comes from the nature of fully polarized reference
state.

There are two particular kinds of currents for which $\delta E$ is easy to
calculate. The first kind is $k_\sigma =0$ for $\sigma \neq \nu$.
Then the generator boosts all particles in the first quantization by the equal
momentum $-k_\nu$, and commutes with $\tilde{P}_{ij}$.
Then we obtain
\begin{equation}
{\cal H}\phi_{\{k\}} \Psi_G = \Psi_G \sum_{i=1}^M p_i^2 \phi_{\{k\}} +
2\sum_{i=1}^M (p_i \phi_{\{k\}}) p_i\Psi_G +\phi_{\{k\}} {\cal H}\Psi_G,
\label{eq:mom}
\end{equation}
where the second term on the right hand side is zero.
The increment of energy is thus given by
$$
\delta E = (N_L-N_\sigma)k_\nu^2 .
$$
The second kind of currents corresponds to the charge current, and is given by
$k_\sigma = k$ for all $\sigma$.  In this case the exponent in $\phi_{\{k\}}
\{\theta\}$ cancels each other except for that of holes.  At this point we
notice that for the singlet state the transfer of electrons with spin $\nu$ is
in fact equivalent to transfer of any kind of spin.  Then for the singlet state
the effective Hamiltonian is given by
\begin{equation}
{\cal H}_{eff} = \nu \sum_{i=1}^Q p_{ih}^2 + \sum_{\sigma\, (\neq \nu)}\sum_i
p_{i\sigma}^2 +{1\over 4}\sum_{i\neq j}\sin ^{-2}({1\over 2} \theta _{ij})(1 +
{P}_{ij}) + E_M,
\label{eq:eff}
\end{equation}
where $P_{ij}$ is the SU($\nu -1$) permutation operator.  Note that the hole
degrees of freedom appear in the first term only.  We emphasize, however, that
the hard-core repulsion between holes and spin bosons must be taken into
account in the solution of eq.(\ref{eq:eff}) because the effective Hamiltonian
is valid only for the lattice model.
With this restriction, we can still proceed in the same way as in
eq.(\ref{eq:mom}) and obtain
$$
\delta E = \nu k^2 Q.
$$

We have thus two independent equations which are enough to determine the spin
and charge velocities $v_s$ and $v_c$.  In the original energy unit they are
calculated as
$$
v_s = v=\pi t,  \ \ \ v_c =u+v=\pi t(1-n_e).
$$
This result is consistent with refs.1 and 6.


\vspace{10mm}
\noindent
{\bf DISCUSSIONS}

\bigskip
In the present derivation of current excitation spectra, it is clear that the
quadratic dependence of $\delta E$ on currents is due the parabolic spectrum of
the model.  This dependence is not restricted to infinitesimal currents as long
as the momentum distribution is within certain limit.$^{1-6}$  The parabolic
spectrum is also responsible for the spin-charge separation for all energies.
Our treatment here is restricted to the singlet ground state from which
currents are generated.  It should be interesting to introduce a generator to
polarize the ground state, and to study the increment of energy.  The generator
may be related to the Yangian discussed in ref.22.

In the momentum space the \it t-J \rm model is written as
\begin{equation}
{\cal H} = -\sum_{k=1}^{N_L} (k-\frac{N_L}{2})^2 \sum_{a,b} X^{ab}(k)
X^{ba}(-k)\theta _b+E_0,
\label{eq:k-rep}
\end{equation}
where $X^{ab}(k)$ is the Fourier transform of $X_i^{ab}$ and $E_0$ is a
constant dependent on the number of vacant sites.  As we have considered
polynomial wave functions of $z_i=\exp(i\theta _i)$ in the first quantization,
we can also restrict operators to polynomials of $X^{ab}_i$.  Under this
restriction $(q-N_L/2)$ in eq.(\ref{eq:fou}) is replaced by
$-i\partial/\partial\theta$ which acts on $X$-operators.  Then the model can
also be written as
\begin{equation}
{\cal H}= -N_L\,\int_0^{2\pi}\frac{{\rm d}\theta}{2\pi}\sum_{a,b}\
\frac{\partial X_\theta^{ab}}{\partial\theta}\cdot\frac{\partial
X_\theta^{ba}}{\partial\theta}
\theta_b+E_0.
\label{eq:sigma}
\end{equation}
Inspired by this form, we have recently introduced a new supersymmetric model
defined in a $d$-dimensional continuum space.$^{16}$  The model is given by
\begin{equation}
{\cal H}={1\over 2}g\int {\rm d}\itvec r \sum\limits_{a, b} \nabla X^{ab}
(\itvec r) \cdot \nabla X^{ba}(\itvec r)\theta_b+C ,
\label{eq:mod}
\end{equation}
where $\nabla =\partial/\partial\itvec r$. In this continuum model one needs to
perform renormalization to remove divergences.  The constants $g$ and $C$ are
determined so as to give zero energy for the vacuum and finite energy for a
finite number of hard-core fermions.  Note that the overall sign in
eq.(\ref{eq:mod}) is reversed from that of eq.(\ref{eq:sigma}).

In ref.16 we have shown that after renormalization all wave functions and
eigenvalues are the same as those of $\nu$-component free fermions.
The presence of supersymmetry is essential in this surprising cancellation of
the hard-core constraint against the exchange-type interaction.  However, there
is no spin-charge separation in this continuum model.

The author would acknowledge useful discussions with N. Kawakami, H. Yokoyama
and J. Zittartz.

\newpage
\noindent
{\bf REFERENCES} \\
\begin{itemize}
\parsep=0ex
\itemsep=0ex
\item[{1.}] Y. Kuramoto and H. Yokoyama, Phys.Rev.Lett. {\bf 67}, 1338
(1991); Physica {\bf C185-189}, 1557
(1991).
\item[{2.}] H. Yokoyama and Y. Kuramoto, J.Phys.Soc.Jpn. {\bf 61},
3046 (1992).
\item[{3.}] Y. Kuramoto, Physica {\bf B186-188}, 831 (1993).
\item[{4.}] N. Kawakami, Phys.Rev. B{\bf 45}, 7525 (1992).
\item[{5.}] D.F. Wang, J.T. Liu and P. Coleman, Phys.Rev. B{\bf 46},
6639 (1992).
\item[{6.}] Z.N.C. Ha and F.D.M. Haldane, Phys.Rev. B{\bf 46}, 9359
(1992).
\item[{7.}] F.D.M. Haldane, Phys.Rev.Lett. {\bf 60}, 635 (1988).
\item[{8.}] B.S. Shastry, Phys.Rev.Lett. {\bf 60}, 639 (1988).
\item[{9.}] A.P. Polychronakos, Phys.Rev.Lett. {\bf 69}, 703 (1992); {\bf 70},
2329 (1993).
\item[{10.}] K. Hikami and M. Wadati, Phys.Lett. A{\bf 173}, 263 (1993).
\item[{11.}] B.S. Shastry and B. Sutherland, Phys.Rev.Lett. {\bf 70}, 4092
(1993); {\bf 71}, 5 (1993).
\item[{12.}] M. Fowler and J.A. Minahan, Phys.Rev.Lett. {\bf 70}, 2325 (1993).
\item[{13.}] D.F. Wang and M. Gruber, preprint.
\item[{14.}] J. A. Minahan and A.P. Polychronakos, Phys.Lett. B{\bf 302}, 265
(1993).
\item[{15.}] K. Vacek, A. Okiji and N. Kawakami, preprint.
\item[{16.}] Y. Kuramoto and J. Zittartz, Phys.Rev.Lett. {\bf 72}, 442 (1994).
\item[{17.}] E. Witten: Commun. Math. Phys. {\bf 92}, 455 (1984).
\item[{18.}] J.F. Cornwell, \it Group Theory in Physics III,
Supersymmetries and Infinite Dimensional Algebras \rm (Academic Press,
1989).
\item[{19.}] P.B. Wiegmann, Phys.Rev.Lett.{\bf 60}, 821 (1988); D. F\"{o}rster,
Phys.Rev.Lett.{\bf 63}, 2140 (1989).
\item[{20.}] B. Sutherland, Phys.Rev. B{\bf 38}, 5589 (1988).
\item[{21.}] B. Sutherland, Phys.Rev. A{\bf 4}, 2019 (1971).
\item[{22.}] F.D.M. Haldane et al., Phys.Rev.Lett. {\bf 69}, 2021 (1992).

\end{itemize}

\end{document}